\documentclass[onecolumn]{svjour2}                    
\smartqed  
\usepackage{graphicx}
\usepackage{bm}
\usepackage{amsmath}
\journalname{Foundations of Physics}
\begin{document}

\title{A novel interpretation of the Klein-Gordon equation}


\author{K.B. Wharton}


\institute{Ken Wharton \at
              Department of Physics and Astronomy\\San Jos\'{e} State University, San Jos\'{e}, CA 95192-0106 \\
              \email{wharton@science.sjsu.edu}
              }

\date{Received: date / Accepted: date}

\maketitle

\begin{abstract}
The covariant Klein-Gordon equation requires twice the boundary conditions of the Schr\"odinger equation and does not have an accepted single-particle interpretation.  Instead of interpreting its solution as a probability wave determined by an initial boundary condition, this paper considers the possibility that the solutions are determined by both an initial and a final boundary condition.  By constructing an invariant joint probability distribution from the size of the solution space, it is shown that the usual measurement probabilities can nearly be recovered in the non-relativistic limit, provided that neither boundary constrains the energy to a precision near $\hbar/t_0$ (where $t_0$ is the time duration between the boundary conditions).  Otherwise, deviations from standard quantum mechanics are predicted.
\keywords{Relativistic Quantum Mechanics \and Classical Fields \and Quantum Foundations}
\PACS{03.65.Pm \and 03.65.Ta}
\end{abstract}

\maketitle

\setlength{\baselineskip}{1.43\baselineskip} 

\vspace{\baselineskip}
\section{Introduction}

Because the Klein-Gordon equation (KGE) for a classical, scalar field was the conceptual precursor to Schr\"odinger's ``derivation'' of the Schr\"odinger Equation (SE), there is a widely-held misconception that the SE must be the non-relativistic limit of the KGE.  In fact, a second-order (in time) differential equation like the KGE does not reduce to a \textit{single} first-order differential equation (like the SE) in \textit{any} limit.\footnote{See section 6 for a related discussion.}  One can reduce the KGE to a single SE only by artificially discarding half of the solutions -- the so-called ``negative-energy'' solutions that evolve like $exp(+i\omega t)$.  This may seem reasonable when considered in the framework of standard, non-relativistic quantum mechanics (NRQM), but NRQM must be a limit of a generally covariant theory where such a step is not natural.  Indeed, this historical division between two arbitrary halves of the solution has created severe problems for quantum gravity, for in curved space-time there is no well-defined way to separate the positive- and negative-energy terms \cite{DeWitt}.

This paper motivates and analyzes a novel interpretation of the KGE (corresponding to a neutral, spinless particle) that promises to reproduce the predictions of standard NRQM \textit{without} arbitrarily discarding any of the solutions.  Efforts to give the KGE a probabilistic interpretation analogous to the SE have historically led to two problems: apparent negative-energy solutions and no positive-definite probability four-current.  Several recent papers have addressed these issues \cite{Horton}\cite{Mostaf}\cite{Kleefeld2}\cite{Nikolic2}, but there is not yet any accepted resolution.  Both of these problems arise from the presumption that any interpretation of the KGE must be directly analogous to the standard interpretation of the SE.  However, such a direct analogy is inherently unlikely, due to the mathematical disconnect between first-order and second-order differential equations.  Specifically, while the full solution to the SE requires only an initial boundary condition $\psi(t=0)$, the full solution to the KGE requires two independent initial boundary conditions: both $\phi(t=0)$ and the first time derivative $\dot{\phi}(t=0)$.  Without some expanded measurement theory to explain how to simultaneously impose independent boundary conditions on both $\phi$ and $\dot{\phi}$, one cannot even solve the KGE, let alone interpret the solutions.  

But if there was some way to impose additional initial boundary conditions (outside the scope of standard QM), then that extra information should allow for experimental predictions that surpass QM.  A century of experimentation has not accomplished this feat, providing strong evidence that one cannot physically impose the necessary initial conditions to solve the KGE.  In other words, for a generic (neutral, spinless) particle, the maximum amount of knowable information can be encoded by the instantaneous values of a single complex scalar field $\psi$ -- only half as much information as the instantaneous values of the $\phi$ and $\dot{\phi}$ fields needed to solve the KGE for a complex scalar field.  This ``half-knowledge'' situation is curiously similar to the axiomatic foundation of Spekkens's interesting toy model \cite{Spekkens}, but still leaves the unanswered question: how can one calculate the solutions to a second-order differential equation like the KGE without sufficient initial boundary conditions?

I propose that a natural resolution to this dilemma can be found by considering time-symmetric approaches to NRQM; namely, those in which one imposes two boundary conditions \textit{at two different times} corresponding to a preparation and a measurement \cite{Aharanov}\cite{Sutherland}\cite{Wharton}.  If only half of the required information to solve the KGE can be imposed as an initial boundary condition, then a subsequent measurement can still impose the remaining constraints as a final boundary condition.  This approach can also be extended to curved space-time by imposing a closed hypersurface boundary condition on classical fields (similar to the closed boundaries imposed on quantum fields in recent work by Oeckl \cite{Oeckl}\cite{Oeckl2}).  

To compare this approach to NRQM, one simply constrains the KGE by imposing two boundary conditions on two \textit{different} instantaneous hypersurfaces (corresponding to consecutive external interactions/measurements).  The solution to the KGE can then be ``retrodicted" in the space-time volume between the two measurements.  Because the full solution cannot be determined by the initial measurement alone, one is forced into a probabilistic description, weighting the solution space to determine the relative likelihood of different pairs of measurements.  After the final measurement, when the additional boundary conditions become known, such a probabilistic interpretation is no longer necessary.  The central result of this paper is to demonstrate the existence of a relativistically invariant weighting scheme that, in the non-relativistic (NR) limit, gives probabilities very similar (but not identical) to NRQM.  Furthermore, the historical problems with the KGE are solved by this approach: all possible solutions to the KGE represent positive energy (as is already the case for a classical scalar field), and the probabilities are always positive definite.

This paper is organized as follows:  Section 2 solves the two-boundary Klein-Gordon equation for generic time-even boundary conditions, circumventing the infinite poles in the same manner as in quantum field theory.  Section 3 then motivates an invariant joint probability distribution, and demonstrates that it is nearly identical to standard conditional probabilities in the NR limit (given other constraints that are derived in an appendix).  Section 4 addresses time-odd, multiple, and incomplete measurements.  Section 5 examines the constraints from the appendix, and finds that they are closely related to a time-energy uncertainty principle.  Section 6 summarizes the postulates used in this approach, details some of the next steps required to extend this research program, and touches on other implications.

\section{Two-Time Boundary Conditions}

In the absence of a potential, the Klein-Gordon equation on a complex scalar field $\phi$ (and its solutions) can be written 
\begin{equation}
	\label{eq:kge}
	\left( \frac{\partial^2}{\partial t^2} -c^2\nabla^2 + \frac{m^2c^4}{\hbar^2}\right)\phi=0,
	\end{equation}
	\begin{equation}
	\label{eq:sols}
	\phi(\bm{r},t)= \int a(\bm{k})e^{i(\bm{k\cdot r}-\omega t)} +  b(\bm{k})e^{i(\bm{k\cdot r}+\omega t)} \bm {dk}.
	\end{equation}
Here $a(\bm{k})$ and $b(\bm{k})$ are complex functions, and $\omega$ is assumed to be a positive function of $\bm{k}$; $\omega (\bm{k})=\sqrt{c^2 k^2+m^2c^4/\hbar^2}$.  

Solutions of the form (\ref{eq:sols}) have exactly twice the independent parameters as the solutions to the Schr\"odinger Equation (for which the $b(\bm{k})$ terms are all identically zero).  The analysis from the previous section can equally well be applied to this doubled parameter space: If an initial preparation constraint can only specify parameters equivalent to $a(\bm{k})$ (a single complex function), then specifying the two independent complex functions in the KGE solutions must require twice the constraints.  Placing a second constraint at the next measurement is independently motivated by CPT-symmetry \cite{Wharton}, but this symmetry implies both boundaries should be mathematically equivalent -- not a strict initial boundary followed by some  discontinuous ``projection" at the final boundary.

There have been several prior efforts to impose two-time boundary conditions in NRQM without doubling the parameter space of the Schr\"odinger equation \cite{Gellmann}\cite{Schulman}; these efforts have revealed that the equations generally become overconstrained.  In light of this difficulty, other two-boundary efforts have used a doubled-parameter space \cite{Aharanov,Sutherland,Cramer}, where the ordinary quantum wavefunction $\psi$ is constrained by an initial boundary and some \textit{other} wavefunction is constrained by a final boundary.  In the present work, no artificial division is made between any two halves of the full Klein-Gordon field.  (Recall, it is this arbitrary splitting into so-called positive-energy and negative-energy solutions that causes trouble in curved space-time.)  Instead, this approach uses only one field, $\phi$, \textit{all} of which is constrained by both an initial and a final boundary.  (Using this same approach, Rovelli has calculated the action of a classical scalar field \cite{Rovelli}, but no direct use of this result has been made except for a noted similarity to the quantum field propagator.)

In order to impose boundary conditions, it is convenient to use conventional measurement theory from NRQM, where a measurement constrains $\phi$ to be an eigenfunction of some operator at a particular time (on a space-like hypersurface $\bm{s}$).  This permits a boundary to be imposed on, say, $\phi(\bm{r},t=0)$, but only in the reference frame somehow defined by the measurement itself.  (This will eventually have to be generalized, as discussed in section 6.)  Another complication is that the momentum operator in physical space $\bm{P}\rightarrow -i\hbar\nabla$  has been constructed for the SE, not the KGE.  Applying it to solutions of the form (\ref{eq:sols}) leads to unphysical results, because the $b(\bm{k})$ terms physically propagate in the opposite direction as determined by the eigenvalues of $\bm{P}$.  This is a general problem for any time-odd operator applied to the KGE.  A natural solution will be presented in section 4; for now the analysis is simply restricted to time-even measurement operators such as $\bm{X}$ and $\bm{P}^2$.

Suppose the eigenfunction of the initial time-even measurement operator on the hypersurface $t=0$ is $f(\bm{r})$, and the eigenfunction of the final time-even measurement operator on the hypersurface $t=t_0$ is $g(\bm{r})$.  (These eigenfunctions are determined via measurements in precisely the same manner as in standard NRQM.)  Fourier-expanding (and ignoring the $2\pi$'s, which will be automatically incorporated into a later normalization) leads to the general boundary conditions 
\begin{equation}
\label{eq:ibc1}
\phi(\bm{r},0)= f(\bm{r}) = \int F(\bm{k})e^{i\bm{k\cdot r}} \bm{dk},
\end{equation}
\begin{equation}
\label{eq:fbc1}
\phi(\bm{r},t_0)= g(\bm{r}) = \int G(\bm{k})e^{i\bm{k\cdot r}} \bm{dk}.
\end{equation}
Comparing these equations to (\ref{eq:sols}), the uniqueness of the Fourier transform implies
\begin{equation}
\label{eq:a1}
a(\bm{k})+b(\bm{k})=F(\bm{k}),
\end{equation}
\begin{equation}
\label{eq:b1}
a(\bm{k})e^{-i\omega t_0}+b(\bm{k})e^{i\omega t_0}=G(\bm{k}).
\end{equation}
This is the correct number of equations to solve for the coefficients $a(\bm{k})$ and $b(\bm{k})$, but there is a problem for the particular values of $\bm{k}$ at which $\omega(\bm{k_}n)t_0=n\pi$.  At these discrete values, solving (\ref{eq:a1}) and (\ref{eq:b1}) yield expressions for $a(\bm{k})$ and $b(\bm{k})$ with infinite poles (for arbitrary boundaries $F$ and $G$).

But quantum field theory is no stranger to infinite poles -- the propagator for the KGE contains a term $(E^2-p^2c^2-m^2c^4)^{-1}$ which blows up on the mass shell.  The standard prescription for solving this problem is to make the assignment $m^2 \rightarrow m^2-i\epsilon$.  After performing the integral, one takes the limit $\epsilon \rightarrow 0$.  Because this is known to give acceptable results, it seems reasonable to use precisely the same method here.  Implementing this change in $m^2$ changes the solutions to the KGE;
	\begin{equation}
	\label{eq:sols2}
	\phi(\bm{r},t)= \int^\infty_{-\infty} a(\bm{k})e^{i(\bm{k\cdot r}-\omega t)}e^{-\epsilon t} +  b(\bm{k})e^{i(\bm{k\cdot r}+\omega t)}e^{\epsilon t} \bm {dk}.
	\end{equation}
where, as usual, all positive constants are absorbed into $\epsilon$ itself.   (A $1/\omega$ term has also been absorbed into $\epsilon$, which is approximately constant in a NR approximation.)  One concern raised by the form of (\ref{eq:sols2}) is that it appears to diverge as $t\rightarrow\pm\infty$.  But recall that we are imposing both an initial and a final boundary condition, so between these boundaries $\phi$ remains finite (as we eventually will take $\epsilon\rightarrow 0$).  

Now it is possible to solve the two boundary problem, using the new form of the solution (\ref{eq:sols2}).  The original boundary conditions (\ref{eq:a1}) and (\ref{eq:b1}) now appear as
\begin{equation}
\label{eq:a2}
a(\bm{k})+b(\bm{k})=F(\bm{k}),
\end{equation}
\begin{equation}
\label{eq:b2}
a(\bm{k})e^{-i\omega t_0}e^{-\epsilon t_0}+b(\bm{k})e^{i\omega t_0}e^{\epsilon t_0}=G(\bm{k}).
\end{equation}
These are now exactly solvable for any two generic boundary conditions, yielding
\begin{equation}
\label{eq:a3}
a(\bm{k})=\frac{F(\bm{k})e^{i\omega t_0}e^{\epsilon t_0}-G(\bm{k})}{e^{i\omega t_0}e^{\epsilon t_0}-e^{-i\omega t_0}e^{-\epsilon t_0}},
\end{equation}
\begin{equation}
\label{eq:b3}
b(\bm{k})=\frac{G(\bm{k})-F(\bm{k})e^{-i\omega t_0}e^{-\epsilon t_0}}{e^{i\omega t_0}e^{\epsilon t_0}-e^{-i\omega t_0}e^{-\epsilon t_0}}.
\end{equation}
Notice that the addition of the small quantity $\epsilon$ now prevents the poles at $\omega(\bm{k_}n)t_0=n\pi$ from diverging.  

Before proceeding, it is worth noting that this technique has already managed the feat of finding a continuous field $\phi(\bm{r},t)$ everywhere between any pair of time-even measurements.  No ``projection'' is needed to get from one exact measurement to the next; one simply uses (\ref{eq:a3}) and (\ref{eq:b3}) to solve (\ref{eq:sols2}) exactly.  Still, this cannot be done until after the second boundary becomes known, so this is a retrodiction, not a prediction.  Also, some of the coefficients still diverge as $\epsilon\to 0$, making it difficult to ascribe any ``reality'' to $\phi$.  Various physical motivations for a non-zero $\epsilon$ will be discussed in section 6.  For now, one can take the already-common ``agnostic'' viewpoint, treating $\phi$ as simply a mathematical tool for calculating measurement probabilities.  In the next section, it is shown that this can be accomplished even if $\epsilon\to 0$ exactly.

\section{Measurement Probabilities}

Given a two-time-boundary framework, the use of conditional probability needs to be revisited.  In standard NRQM, some initial measurement $(\bm{Q}_F,F)$ at time $t\!\!=\!\!0$ determines the state/wavefunction $\psi(t\!=\!0)$.  (In this notation, $\bm{Q}_F$ is the experimenter's choice of measurement, and $F$ is the particular outcome.)  Evolving $\psi$ to the time $t_0$ of the next measurement $(\bm{Q}_G,G)$, one traditionally uses the conditional probability $P(G|\psi(t_0),\bm{Q}_G)$.  But this time-directed process simply does not work for the KGE because $(\bm{Q}_F,F)$ is not sufficient to determine the solution $\phi$.  Furthermore, this paper speculates that $(\bm{Q}_G,G)$ determines $\phi(t_0)$, so one could hardly have a conditional probability of $G$ in turn determined by $\phi(t_0)$.  

In this framework, one needs a time-symmetric implementation of probability, closer in spirit to scattering amplitudes in quantum field theory.  Such a symmetry can be naturally achieved by treating joint probability distributions (JPDs) as the fundamental quantities from which conditional probabilities can be derived.  This approach assigns a joint probability to every possible initial/final measurement outcome pair $F_i,G_j$ and the time between them $\Delta t=t_0$.  If there was some JPD given by $J(F_i,G_j,\Delta t)$ associated with each possible pair of outcomes (for a given pair of measurement operators $\bm{Q}_F$, $\bm{Q}_G$) one could generate any particular conditional probability $P(G_1|F_2,\Delta t)$ via the standard normalization procedure: $J(F_2,G_1,\Delta t)/\sum_jJ(F_2,G_j,\Delta t)$.  For operators with continuous eigenvalues, the sum becomes an integral over all eigenstates of the chosen operator $\bm{Q}_G$.  Standard zero-potential probabilities in NRQM can be recovered from the JPD
\begin{equation}
\label{eq:Po}
J_0(F,G,t_0)=\left| \int F(\bm{k})G^*(\bm{k})e^{-i\omega t_0}\bm{dk}\right|^2.
\end{equation}

For any given $F$ and $t_0$, the conditional probability of measuring any function $G$ is given by this same expression, $P_0(G|F,t_0)\!=\!J_0(F,G,t_0)$, if one constrains the magnitude of $F$ and $G$ such that $\int |F|^2 \bm{dk}=\int |G|^2 \bm{dk}=1$.  (Note these constraints are simply chosen for familiarity, and should not be interpreted as a separate normalization; changing the ``1'' to any constant would have no effect, due to the normalization in the standard $J\to P$ procedure described in the previous paragraph.)  Of course, the above expression for $J_0$ is not relativistically invariant.  The challenge for any relativistic extension of quantum mechanics is to devise an invariant JPD that is equal to $J_0$ in the appropriate limit.

Note that JPDs are not probability densities, so the conditional probabilities derived from them need not satisfy a continuity equation.  Indeed, to talk about the probabilities of a measurement at a time between $t=0$ and $t=t_0$ is a contradiction in terms, because the next measurement happens at $t=t_0$ \textit{by definition}.  A similar conclusion was reached in Oeckl's recent work on ``general boundary'' quantum field theory \cite{Oeckl}.  

In relativistic quantum mechanics, the charge density $\rho(\bm{r},t)$ cannot represent probability because it can be negative.  Still, because of the usefulness of this density in relativistic quantum mechanics, this is a reasonable place to begin searching for an appropriate JPD.  For the Klein-Gordon equation, the charge density is given by
\begin{equation}
\label{eq:rho}
\rho(\bm{r},t)=\frac{i\hbar}{2mc^2}\left( \phi^* \frac{\partial \phi}{\partial t}-\phi \frac{\partial \phi^*}{\partial t}\right),
\end{equation}
and is real, although not invariant (it is one component of a four-current).  It is not possible to make $\rho$ invariant without some reference unit four-vector that defines the time direction.  

Fortunately, the boundary conditions define two hypersurfaces, $\bm{s}_1$ and $\bm{s}_2$.  These surfaces have associated inward-pointing normal four-vectors $\eta_1$ and $\eta_2$ (pointing forward in time from the initial boundary and backward in time from the final boundary).  In the limit comparable to NRQM, $\bm{s}_1$ and $\bm{s}_2$ are the hyperplanes $t\!=\!0$ and $t\!=\!t_0$, so it is plausible that $\eta$ is the needed reference unit vector, but of course it is only defined on the hypersurfaces.  Because of this constraint, $\rho$ is not well-defined in the volume \textit{between} the boundaries -- but one can still integrate $\rho$ along the hypersurfaces themselves.  So, generalizing to any closed hypersurface defined by the two boundary conditions (e.g. the two infinite planes $t=0$ and $t=t_0$), the hypersurface integral
\begin{equation}
\label{eq:W1}
W=\frac{\hbar}{mc}\oint_{\bm{s_1,s_2}} Im(\phi \eta^\mu \partial_\mu \phi^*) \bm{ds}
\end{equation}
is a scalar that is plausibly related to probability, motivated both by invariance and known results from relativistic quantum mechanics.  (In this notation, $\eta^\mu$ is the inward-pointing four-vector unit normal to the integration surface and $\partial_\mu$ is the four-gradient with $\partial_0=\partial/\partial(ct)$.  Summation over the index $\mu$ is implied.)

In the special case that the closed hypersurface is defined by the planes $t=0$ and $t=t_0$, $W$ can be evaluated using (\ref{eq:sols2}).  (Expand both $\phi$ and $\phi^*$ as integrals in $\bm{k}$ and $\bm{k^\prime}$, and then the spatial integral yields $2\pi\delta(\bm{k}-\bm{k^\prime})$, leaving only a single integral in $\bm{k}$.)  Dropping the overall constants, the simplified result is
\begin{equation}
\label{eq:W2}
W=\int \omega |a|^2\left(1-e^{-2\epsilon t_0}\right)-\omega|b|^2\left(1-e^{2\epsilon t_0}\right) -Im\left[2\epsilon ab^*\left(1-e^{-2i\omega t_0}\right)\right] \bm{dk} .
\end{equation} 
Next, $a(\bm{k})$ and $b(\bm{k})$ can be written in terms of the boundary conditions $F(\bm{k})$ and $G(\bm{k})$ using (\ref{eq:a3}) and (\ref{eq:b3}).  To lowest surviving orders in $\epsilon$, this yields
\begin{equation}
\label{eq:W3}
W=\int \frac{\omega \epsilon t_0\left[ |F|^2+|G|^2\right]-2\omega\epsilon t_0Re[FG^*]cos(\omega t_0)+\epsilon sin(\omega t_0)(...)}{\epsilon^2 t^2_0 +sin^2(\omega t_0)}  \bm{dk} .
\end{equation}
This does not vanish as  $\epsilon\to 0$ because one gets a periodic delta function from the identity
\begin{equation}
\label{eq:limit}
\lim_{\epsilon\to 0}\,\frac{\epsilon t_0}{\epsilon^2 t^2_0 +sin^2(\omega t_0)}=\sum_{n=0}^\infty \pi\,\delta(\omega t_0-n\pi).
\end{equation}
The $sin(\omega t_0)$ term in the numerator of (\ref{eq:W3}) was not fully shown because it is odd over the delta function, and vanishes; the $cos(\omega t_0)$ term is either $1$ or $-1$ depending on $n$.  Expanding $\bm{dk}$ in spherical $\bm{k}$-coordinates ($k=|\bm{k}|$), one can use the relationship $k\,dk=\omega\, d\omega$ to reduce the radial integral to a sum, leaving only a 2-D integral;
\begin{equation}
\label{eq:W4}
W= \int \sum_{n=n_0}^\infty \frac{-1^n\omega^2_nk_n}{t_0}\left( |F_n|^2+|G_n|^2-2Re[F_nG^*_n] \right)  sin\theta_k  d\theta_k \, d\phi_k.
\end{equation}
In this notation $F_n=F(\bm{k_n})$, $G_n=G(\bm{k_n})$, and $k_n=|\bm{k_n}|$, where $\bm{k_n}$ is the solution to $\omega_n(\bm{k_n})=n\pi /t_0$, and $n_0$ is the smallest value of $n$ for which there is such a solution.

I now propose that an invariant joint probability distribution can be constructed from $W$ according to
\begin{equation}
\label{eq:prob}
J=(W_{max}-W_{min})^2.
\end{equation}
Here $W_{max}$ is the maximum value $W$ can attain when varying any unconstrained parameters; similarly $W_{min}$ is the minimum value.  Such a range parameter is common in probability theory: one way to interpret (\ref{eq:prob}) is that any pair of boundary conditions permits a two-dimensional space of solutions, with the allowed range in each dimension determined by allowed values of $W$.  Picking a solution at random, particular pairs of boundaries that have a larger solution space would be more likely than other pairs with a smaller solution space.

Still, $W$ has no range at all if the parameters $F$, $G$, and $t_0$ are all exactly specified.  But regardless of the precision of the measurements, standard measurement theory always allows for one free parameter: the unknown relative phase between the two eigenfunctions $f(\bm{r})$ and $g(\bm{r})$.  This relative overall phase, $\theta$, does not affect the first two terms in (\ref{eq:W4}), so for complete measurements these terms will not contribute to any calculation of $J$ using (\ref{eq:prob}). 

$J$ must match the standard JPD in the NR limit (\ref{eq:Po}), but $J_0$ is given by a 3-D integral over k-space, not a 2-D integral and a discrete sum.  Still, the sum in (\ref{eq:W4}) approximates an integral if neither $F(\bm{k})$ nor $G(\bm{k})$ change very rapidly when the magnitude $|\bm{k}|$ changes by $k_{n+1}-k_n$; the precise conditions are derived in the appendix and discussed in section 5.  Given these constraints, the appendix demonstrates that the third term in (\ref{eq:W4}) can be excellently approximated as
\begin{equation}
\label{eq:approx}
\int \!\! \sum_{n=n_0}^\infty \!\! \frac{-1^nk_n}{t_0}Re[F_nG^*_n] sin\theta_k  d\theta_k \, d\phi_k\approx \frac{\hbar}{2m}\!\int \!\! Re[F(\bm{k})G^*(\bm{k})]cos(\omega t_0) \bm{dk}.
\end{equation}

To examine whether or not $J\approx J_0$, one must now use the fact that both $f(\bm{r})$ and $g(\bm{r})$ were originally eigenfunctions of time-even operators (time-odd operators are considered in the next section).  The eigenfunctions of non-degenerate time-even operators can be constrained to be entirely real if their relative phase $exp(i\theta)$ is added explicitly \cite{Sakurai}.  (For the degenerate case, there is no relative phase information, so one is still free to choose real combinations of the real eigenfunctions.)  So, forcing $f(\bm{r})$ and $g(\bm{r})$ to be real by adding the relative overall phase $exp(i\theta)$, the approximation (\ref{eq:approx}) simplifies the JPD (\ref{eq:prob}) into the expression
\begin{equation}
\label{eq:prob2}
J=\left( \int Re[F(\bm{k})G^*(\bm{k})e^{i\theta}]cos(\omega t_0)\,\,\bm{dk} \right)^2_{max}.
\end{equation}
This assumes that one is in the NR limit such that the $\omega^2_n$ in (\ref{eq:W4}) is roughly constant and can be pulled out of the integral (overall constants are irrelevant to the unnormalized JPD).  Given the complete freedom of the unmeasurable quantity $\theta$, (\ref{eq:prob2}) implies $W_{max}=-W_{min}$, so the range $W_{max}-W_{min}$ is just $2 W_{max}$.  

Additional simplifications result from $f(\bm{r})$ and $g(\bm{r})$ being real; both $Im[F(k)G^*(k)]cos(\omega t_0)$ and $Im[F(k)G^*(k)]sin(\omega t_0)$ are odd functions in $k$, so their integrals cancel out of both $J_0$ (\ref{eq:Po}) and $J$ (\ref{eq:prob2}).  Extracting the k-independent portion $\omega_0=mc^2/\hbar$ from $\omega(k)=\omega_0+\omega_1(k)$ further simplifies these expressions to
\begin{equation}
\label{eq:Po2}
J_0=\left( \int Re[FG^*]cos(\omega_1 t_0)\,\,\bm{dk} \right)^2+\left( \int Re[FG^*]sin(\omega_1 t_0)\,\,\bm{dk} \right)^2,
\end{equation}
\begin{equation}
\label{eq:prob3}
J=\left( \int Re[FG^*]cos\theta\left[cos(\omega_0 t_0)cos(\omega_1 t_0)-sin(\omega_0 t_0)sin(\omega_1 t_0)\right]\bm{dk} \right)^2_{max}.
\end{equation}

While similar, these two expressions are unfortunately not equivalent if $t_0$ is constrained to be a precise value.  But is practically impossible to constrain $t_0$ to an accuracy less than $\omega^{-1}_0$ ($\sim \!\!10^{-20}$ seconds for an electron, and even less for more massive particles).  No current experiments can determine the measurement time to a sub-attosecond accuracy, so $W$ must be maximized over the completely unknown angle $\omega_0t_0$ as well as over $\theta$.  The latter maximization is trivial, as one simply forces $cos\,\theta=1$ in (\ref{eq:prob3}).  And with complete freedom of the angle $\omega_0t_0$, the maximum value of $(A\,sin\omega_0t_0 \pm B\,cos\omega_0t_0)^2$ is just $A^2+B^2$.  Using this fact, one can compare (\ref{eq:Po2}) and (\ref{eq:prob3}) to confirm that $J\approx J_0$, to within the accuracy of the previous approximations.

Note that this procedure only works if the precise value of $t_0$ becomes part of the solution parameter space, and is not merely an unknown random quantity.  In other words, if $t_0$ is constrained to be some particular value -- even if that value is completely unknown -- then $J$ is no longer necessarily equal to $J_0$.  

From a foundational perspective, using $J$ is preferable to $J_0$, because $J$ is relativistically invariant, unlike (\ref{eq:Po}).  But theoretical preference is no longer the only issue; even given realistic uncertainties in $t_0$, the above approximations ensure that $J$ and $J_0$ are \textit{not} exactly the same, opening the door to an experimental differentiation between this approach and NRQM.  $J$ and $J_0$ begin to diverge in the relativistic regime, and also when the approximation (\ref{eq:approx}) begins to fail as discussed in section 5.

The other regime in which new effects would be expected is if the value of $t_0$ could be experimentally constrained to a value much less than $\omega_0^{-1}$.  Although such capabilities currently seem far out of reach, it is worth considering a future experiment that might constrain $\omega_0t_0$ to be approximately a multiple of $\pi$.  In that case, the $sin(\omega_0t_0)$ term in (\ref{eq:prob3}) can be ignored, and one finds that $J$ reduces to only the first term in (\ref{eq:Po2}).  This term is the square the real portion of the standard NRQM transition amplitude, and of course there are many deviations from NRQM that would occur if the imaginary part of the amplitude was ignored.

\section{Time-Odd, Multiple, and Incomplete Measurements}

The previous sections have assumed two consecutive measurements of time-even quantities, such as position, $\bm{P}^2$, or $\bm{L}^2$ (angular momentum).  Measurements of time-odd quantities leads to difficulties, because (as noted in Section 2) the $b(\bm{k})$ terms in (\ref{eq:sols}) physically propagate in the opposite direction as determined by the eigenvalues of $\bm{P}$.  Another problem, as noted in Section 1, is that there are two independent (but equally important) mathematical objects on which initial values can be specified: both $\phi$ and the first time derivative $\dot{\phi}$.  Initial boundaries have historically only been imposed on the field itself because of the reliance on first-order differential equations, but when using the KGE one should consider the possibility of constraining the value of $\dot{\phi}$ as well as $\phi$.

Both of the issues in the previous paragraph are linked to a simple fact: the measurement of a time-even quantity must yield the same value under time-reversal of the entire system, while the measurement of a time-odd quantity (like momentum) must change sign under time-reversal.  In spinless NRQM, time-reversal is accomplished via a complex conjugation, which is why (in the position basis) time-even operators are real and time-odd operators are imaginary.  But complex conjugation does not time-reverse the KGE, only the SE, so using ``$i$'s" to distinguish time-even and time-odd operators must now fail.  What \textit{does} get a sign-change upon time reversal is $\dot{\phi}$, the very object that standard measurement theory does not address.  It is therefore tempting to replace the ``$i$'' in time-odd measurement operators with $-{\partial}/{\partial t}$.  While the operator units are now wrong, this is easily fixed by multiplying these time-odd operators by the natural unit of time $\omega_0^{-1}=\hbar/(mc^2)$.   This section will demonstrate that such a replacement solves the momentum eigenvalue problem while also allowing measurements to constrain $\dot{\phi}$ instead of $\phi$.

Under the substitution $i\to-\omega_0^{-1}\partial/\partial t$ in all time-odd operators, the corresponding eigenvalue equation $\bm{Q}\phi=q\phi$ then changes to become
\begin{equation}
\label{eq:qodd}
i\bm{Q}\dot{\phi}=\omega_0 q \phi
\end{equation}
for any standard time-odd operator $\bm{Q}$ (and eigenvalue $q$ corresponding to eigenfunction $\phi$).

In the limit where one expects the ordinary NRQM operators to be valid, the frequency $\omega(\bm{k})$ at the relevant values of $\bm{k}$ are all approximately the same; $\omega\approx\omega_0=mc^2/\hbar$.  Using this approximation, the time-derivative of (\ref{eq:sols}) is simply
\begin{equation}
\label{eq:phidot}
\dot{\phi}=\frac{\partial\phi}{\partial t}\approx-i\omega_0\int a(\bm{k})e^{i(\bm{k\cdot r}-\omega t)} -  b(\bm{k})e^{i(\bm{k\cdot r}+\omega t)} \bm {dk}.
\end{equation}
Inserting this into (\ref{eq:qodd}), one finds that the ordinary eigenvalue equation $\bm{Q}\phi=q\phi$ is recovered when $b(\bm{k})=0$ (the standard NRQM condition).  On the other hand, if $a(\bm{k})=0$, the equation now has the opposite eigenvalue, $-q$.  This is exactly the sign-reversal needed to solve the momentum eigenvalue problem noted above; if $\bm{Q}$ is $-i\hbar\nabla$, and yields a measurement eigenvalue $\hbar\bm{k}_0$ at $t=0$, then (\ref{eq:qodd}) implies a boundary condition $a(\bm{k})+b(\bm{-k})=\delta(\bm{k}-\bm{k}_0)$.  Regardless of the relative phase or weight between $a(\bm{k}_0)$ and $b(-\bm{k}_0)$, the combination $a(\bm{k}_0)exp(i\bm{k}_0\cdot\bm{r}-i\omega t)+b(-\bm{k}_0)exp(-i\bm{k}_0\cdot\bm{r}+i\omega t)$ physically propagates in a direction aligned with $\bm{k}_0$.  (This would not have been the case if one was using the ordinary eigenvalue equation; the $b(\bm{k}_0)$ term has a phase velocity opposite that of  the $b(-\bm{k}_0)$ term.)

In general, (\ref{eq:qodd}) implies that the measured real eigenfunction (in k-space) $F(\bm{k})$ of a time-odd operator from standard NRQM should be imposed on the combination $a(\bm{k})exp(-i\omega t)+b(-\bm{k})exp(i\omega t)$.  The section 3 conclusion that $J\approx J_0$ therefore remains valid for two consecutive time-odd measurements; one need only change $b(\bm{k})\to b(-\bm{k})$ starting at (\ref{eq:a2}).

It remains to show that a time-odd measurement followed by a time-even measurement (or vice-versa) also gives results consistent with NRQM.  Suppose a time-odd measurement at $t\!=\!0$ yields a k-space measurement with (real) eigenfunction $F(\bm{k})$, and a time-even measurement at $t=t_0$ yields a measurement with (real) eigenfunction $g(\bm{r})$, both within an arbitrary overall phase.  Then the new boundary conditions can be written in terms of $F(\bm{k})$ and $G(\bm{k})$ (the latter is calculated from $g(\bm{r})$ according to (\ref{eq:fbc1})) as
\begin{equation}
\label{eq:a4}
a(\bm{k})+b(\bm{-k})=F(\bm{k}),
\end{equation}
\begin{equation}
\label{eq:b4}
a(\bm{k})e^{-i\omega t_0}e^{-\epsilon t_0}+b(\bm{k})e^{i\omega t_0}e^{\epsilon t_0}=G(\bm{k}).
\end{equation}
These equations have a solution different from (\ref{eq:a3}) and (\ref{eq:b3}).  Substantial but straightforward algebra yields
\begin{equation}
\label{eq:a5}
a(\bm{k})=\frac{F(\bm{k})e^{2i\omega t_0}e^{2\epsilon t_0}+F(\bm{-k})-G(\bm{k})e^{-i\omega t_0-\epsilon t_0}-G(\bm{-k})e^{i\omega t_0+\epsilon t_0}}{e^{2i\omega t_0+2\epsilon t_0}-e^{-2i\omega t_0-2\epsilon t_0}},
\end{equation}
\begin{equation}
\label{eq:b5}
b(\bm{k})=\frac{G(\bm{k})e^{i\omega t_0+\epsilon t_0}+G(\bm{-k})e^{-i\omega t_0-\epsilon t_0}-F(\bm{k})-F(\bm{-k})e^{-2i\omega t_0-2\epsilon t_0}}{e^{2i\omega t_0+2\epsilon t_0}-e^{-2i\omega t_0-2\epsilon t_0}}.
\end{equation}
Inserting these new values of $a(\bm{k})$ and $b(\bm{k})$ into (\ref{eq:W2}) yields a different value of $W$ than given in (\ref{eq:W3}).  The only relevant terms are those with both $F$ and $G$, because all of the terms like $G(\bm{k})G^*(-\bm{k})$ are independent of the relative overall phase $\theta$ and do not contribute to variation in $W$.  Putting in the relative phase explicitly ($FG^*\to exp(i\theta)FG^*$) one can simplify $W$ using the fact that for the real function in k-space $F^*(\bm{k})=F(\bm{k})$ and for the real function in position space $G^*(\bm{k})=G(-\bm{k})$.  To largest surviving orders of $\epsilon$, this yields
\begin{equation}
\label{eq:newW}
W=\int \frac{-\omega \epsilon t_0 Re\left[ F(\bm{k})G^*(\bm{k})\left( e^{3i\omega t_0+i\theta}+e^{-i\omega t_0 + i \theta}+2e^{-i\omega t_0-i\theta}\right) \right]}{\epsilon^2 t^2_0 +sin^2(2\omega t_0)}  \bm{dk} .
\end{equation}

As $\epsilon\to 0$, a periodic delta function arises according to (\ref{eq:limit}), but with twice as many poles as before: $\delta(\omega t_0-n\pi/2)$.  At each pole, $exp(3i\omega t_0)=exp(-i\omega t_0)$, but the value of this quantity picks up a factor of $i$ at each consecutive pole, making both the $Re(FG^*)$ terms and the $Im(FG^*)$ terms important.  Applying the same approximation (\ref{eq:approx}) to both of these terms, one finds that in the NR limit
\begin{equation}
\label{eq:newW2}
W\approx cos\theta\int  Re[F(\bm{k})G^*(\bm{k})]cos(\omega t_0)-Im[F(\bm{k})G^*(\bm{k})]sin(\omega t_0) \, \bm{dk} .
\end{equation}
Using the same procedure as before (pulling out the k-independent part, $\omega(k)=\omega_0+\omega_1(k)$, and varying both the unknown angles $\theta$ and $\omega_0 t_0$), one finds
\begin{equation}
\label{eq:newP}
J\approx \left(Re\!\int \!F(\bm{k})G^*(\bm{k})e^{-i\omega_1 t_0} \bm{dk} \right) ^2 +  \left(Im\!\int \!F(\bm{k})G^*(\bm{k})e^{-i\omega_1 t_0} \bm{dk} \right) ^2\!\!.
\end{equation}
Again, this is equal to the standard JPD $J_0$ -- this time without dropping any terms in $J_0$ due to real eigenfunctions.  Therefore, the analysis from the previous section continues to hold for time-odd measurements followed by time-even measurements (and vice-versa, due to the overall time-symmetry of this approach).

For more than two consecutive measurements, this problem can appear to be over-constrained.  For example, consider a time-even measurement $\bm{Q}_1$ at $t=t_1$, a time-even measurement $\bm{Q}_2$ at $t=t_2$, and a  third time-even measurement $\bm{Q}_3$ at $t=t_3$.  While the outcomes of $\bm{Q}_1$ and $\bm{Q}_2$ fully determine the solution from $t_1\le t \le t_2$, the outcomes of $\bm{Q}_2$ and $\bm{Q}_3$ independently determine the solution from $t_2\le t \le t_3$.  Although this implies a discontinuity in the solution to the KGE at $t=t_2$, note that it is \textit{not} a discontinuity in $\phi$ itself, which is specified by the time-even measurement $\bm{Q}_2$.  Instead, the discontinutity must occur at the unconstrained $\dot{\phi}(t=t_2)$. 

The sort of discontinuity described in the previous paragraph is neither surprising nor unphysical.  If $\phi$ is supposed to describe a quantum system, one can only ``measure'' such a system by interacting with it.  Because the interaction is outside the scope of the KGE, one can hardly hope to find a continuous solution without taking this interaction into account.  Furthermore, the above discontinuity is precisely where the interaction occurs, at $t=t_2$.  If no external system is present at the $t=t_2$ hypersurface, then no information is available on this boundary, and the only external constraints on the system from $t_1\le t \le t_3$ are the measurements at $t=t_1$ and $t=t_3$.  In other words, one effectively ``joins'' together two adjacent regions of space-time into a single region -- precisely as Hardy has recently argued must occur in any eventual theory of quantum gravity \cite{Hardy}.

Incomplete measurements, however, pose a mathematical challenge for this two-boundary approach.  A measurement that does not constrain a complete initial eigenfunction $f(\bm{r})$ has additional ``free'' parameters which must be varied when calculating $W_{max}$ and $W_{min}$.  Possibly the biggest challenge is to determine how to implement commuting incomplete measurements, especially when one measurement is time-even and the other is time-odd (e.g. $\bm{L}^2$ and $\bm{L}_z$).  The key will be to make sure that any discontinuity at the second measurement does not erase the information imposed at the prior (commuting) measurement.  Due to the additional mathematical complexity, this paper can only conjecture that commuting incomplete measurements in this formalism will give results (nearly) equivalent to NRQM.  The most promising path forward seems to be to ignore the time duration between two consecutive commuting measurements, and impose them as simultaneous (but incomplete) constraints on both $\phi$ and $\dot{\phi}$.  

While much work is required to extend this approach to cover the full range of measurement theory from NRQM, such effort may be misguided when it comes to a possible extension to quantum gravity.  That is because the restriction of operators to spacelike hypersurfaces will have to be revisited in a general relativistic framework, possibly requiring a new theory of quantum measurement without the use of operators at all (as in \cite{Wharton3}).  The importance of imposing boundary conditions on arbitrary hypersurfaces will be discussed in section 6. 

\section{A Time-Energy Uncertainty Relation}

When examining the limits in which the previous analysis fails, one finds that some of these limits would manifest themselves as a time-energy uncertainty relation.  Before addressing this result, it should be stressed that the usual position-momentum uncertainty relations are unchanged.  As in standard NRQM, a very precise measurement of $\phi$ in position space leads to a large spread of $\phi$ in k-space due to the Fourier transform (\ref{eq:ibc1}).  There is one difference; unlike standard NRQM, after a pair of complete measurements this approach can now reconstruct $\phi$ between the two measurements, leaving no uncertainty at all (except for the free angles $\theta$ and $\omega_0 t_0$).  Still, this is true classically as well; one can ``beat" Heisenberg's uncertainty principle in hindsight, by measuring the position very accurately at two successive times, and then reconstructing the intermediate velocity.  So knowing a more complete solution to $\phi$ after the fact does not contradict the usual uncertainty principle, which only concerns what can be known at any given time.

In NRQM,  the \textit{time-energy} uncertainty principle has never been put on an even footing with the position-momentum uncertainty principle \cite{Nikolic}, although in a relativistic theory they must of course be intimately related.  The issue is not whether one can measure an arbitrarily precise energy at an arbitrarily precise external clock time -- in principle, one can do that in both this approach as well as in the standard formulation of quantum mechanics \cite{AnB}.  Here, the question becomes whether or not precise energy measurements are reproducible over short periods of time. 

Without using the approximation (\ref{eq:approx}), the exact invariant JPD (\ref{eq:prob}) for two (time-even) energy measurements can be written
\begin{equation}
\label{eq:probexact}
J=\left( \int^\infty_{-\infty} \sum_{n=n_0}^\infty \omega Re[F(\bm{k})G^*(\bm{k})]cos(\omega t_0)\delta(\omega t_0-n\pi)\,\,d^3k \right)^2_{max}.
\end{equation}
From (\ref{eq:probexact}) it is clear that if both measurements are very precise energy measurements, then they must be nearly the same energy (with the same $k$), or else the product $FG^*$ would be very small, leading to a small probability.  But because of the delta function, there is now another constraint; the precisely measured value of $k$ must match up with a particular value $k_n$ (where $k_n$ solves $\omega_n({k_n})=n\pi /t_0$).  This can only happen if $t_0$ is not exactly fixed, but is a parameter that has some freedom to maximize $W$, as discussed at the end of section 3.  This constraint on possible measurement times would never be directly noticeable from a single measurement because of the small changes in $t_0$ ($< 10^{-20}$ sec) needed to set $k=k_n$.

But even with this freedom in the precise value of $t_0$, deviations from standard quantum mechanics will still occur if one cannot  ``ignore" the delta function in (\ref{eq:probexact}), as implied by (\ref{eq:approx}).  And (\ref{eq:approx}) is only correct in certain limits.  The first limit derived in the appendix is (\ref{eq:lim1}), which constrains the rate of change of both $F(k)$ and $G(k)$ for small values of k.  (Specifically, for $k\le k_0$, where $k_0\equiv(2\pi m/\hbar t_0)^{1/2}$.)  For an example of such a constraint, consider a gaussian initial measurement $F(k)=exp[-k^2/(2\sigma^2)]$.  From (\ref{eq:lim1}), one finds that this corresponds to $\sigma>k_0$, equivalent to a spread in the initial energy greater than $h/t_0$.  As this constraint is approached, (\ref{eq:probexact}) predicts that new physics will begin to emerge.  The form of this new physics will be a failure of the ordinary probability distributions, forcing sequential energy measurements give slightly different results.  

Curiously, this ``new physics'' is almost what is expected to happen if there is a time-energy uncertainty relation that is analogous to position-momentum uncertainty.  The difference is that here the time $t_0$ is not an uncertainty in a measurement time, but instead the duration between measurements.  Still, in practice this might be the same thing, as it's not possible to make two measurements (in a well-defined order) separated by a duration smaller than the amount of time it takes to make the measurements. 

The other constraint derived in the appendix also concerns the rate of change of $F(k)$ and $G(k)$; see equation (\ref{eq:lim2}).  For values of $k\approx k_0$, this is a similar constraint to the previously discussed condition (\ref{eq:lim1}).  For larger values of $k$, the condition becomes even less stringent, even for a precise non-zero energy measurement.  Of course, values of $k\ll k_0$ can easily violate this condition, but so long as the constraint (\ref{eq:lim1}) continues to hold, this will just be a very small perturbation on the approximation (\ref{eq:approx}), because the integral on the right side scales like $k^2$, and (\ref{eq:lim1}) forces the important part of the integral to extend out to at least $k=k_0$.

The conclusion is that there are many pairs of boundary conditions for which the approximation (\ref{eq:approx}) does not hold good, leading to a divergence of $J$ from $J_0$, but these pairs of boundaries also approach the time-energy uncertainty limit that is known to be experimentally difficult.  Still, it is promising that these results yield a time-energy uncertainty relation consistent with known experimental limitations.  

\section{Summary, Extensions, and Implications}

Although the results in this paper have not yet been generalized to the point where this is a full alternative interpretation of quantum mechanics, the indications are promising that this approach might put quantum mechanics on a footing consistent with general relativity while also addressing many open interpretational questions.  More immediately, the central result of an invariant scalar from which one can calculate positive definite outcome probabilities for single-particle solutions to the Klein-Gordon equation is itself sufficient motivation to further consider this novel framework.

Three of the four postulates used in the previous sections appear compatible with a general relativistic framework.  These are as follows:

1) The correct wave equation governing generic (spinless, neutral) particles is the Klein-Gordon equation on a complex scalar field $\phi$.  (This is assumed to be a classical field, not the operator-valued field of quantum field theory.)

Although this is a postulate, it still requires some justification.  As noted in the introduction, the KGE and SE are not equivalent in the NR limit.  This can be seen in both the amount of necessary Cauchy data (solving the KGE requires twice the initial data as the SE) as well as in the doubled size of the KGE solution space (see Section 2).  From this perspective, the KGE on a \textit{real} scalar field would be more comparable to the SE, but such a field has no natural U(1) symmetry.  While it is true that the solutions to the complex KGE in the NR limit can always be written as the sum of a solution to the SE and an independent solution to the complex conjugate of the SE, there are an infinite number of other ways to split the KGE solutions into the sum of two arbitrary sub-components.  Assigning physical meaning to one of these particular sub-components under one particular splitting is therefore difficult to justify.  For one, it raises the question of why one can ignore the other term; both components have positive energy, as determined by the zero-potential Hamiltonian $\bm{H}=\bm{P}^2/(2m)$ (as well as by the energy density $T_{00}$ of classical scalar fields).  Ignoring one term also raises severe problems in curved spacetime where there is no covariant splitting of the full solution into two analogous sub-components \cite{DeWitt}.  Therefore it seems more natural to interpret the entire KGE solution as a single entity; that is the goal of this paper.

The next two postulates are:

2) The solution $\phi$ is constrained by a boundary condition on a closed hypersurface $\bm{s}$ (with unit normal four-vector $\eta^\mu$), imposed by external measurements.  Any infinities are dealt with by giving the mass a small imaginary value $\epsilon$, and then taking the limit $\epsilon\to 0$.

3) Each possible closed boundary condition is assigned a joint probability distribution $J=(\Delta W)^2$, where $\Delta W$ is the allowed range of the real invariant scalar $W=\oint_{\bm{s}} Im(\phi \eta^\mu \partial_\mu \phi^*) \bm{ds}$ consistent with the boundary constraints.

The usual conditional probabilities can be generated from $J$ via normalization.  In order to compare the above postulates to a preparation-measurement sequence in standard NRQM, one takes the hypersurface to be the pair of hyperplanes $t\!=\!0$ and $t\!=\!t_0$, presumably connected at some sufficiently large distance where the integrand in W goes to zero.  Imposing boundaries on such hypersurfaces requires a fourth and final postulate;

4) The measured eigenfunction of a time-even operator (from standard NRQM) is imposed on $\phi$ along an instantaneous hypersurface.  Time-odd operators are first adjusted by replacing the ``$i$'' with $-\omega_0^{-1}\partial/\partial t$.

But this use of eigenfunctions and instantaneous surfaces is the only remaining tool from NRQM, and this provides a strong motivation to develop a measurement theory that could apply to \textit{any} hypersurface.  Postulate \#2 implies that any closed hypersurface will suffice, even if portions of that surface are time-like.  Time-like boundary conditions are used in NRQM (e.g. the infinite square well), but are not imposed in the same way as measurements.  This artificial distinction clearly needs to change if postulate \#2 is correct; after all, in general relativity it is not clear whether a given surface is time-like or space-like until the field equations have been solved.  Oeckl and Hardy have recently pointed this out, implying that any candidate theory of quantum gravity must be able to deal with regions bounded by time-like hypersurfaces \cite{Oeckl}\cite{Hardy}.  Such a generalization of measurement theory might also help to address some of the unanswered questions from section 4.  

The framework in this paper falls short of a full alternative interpretation of quantum mechanics because the above analysis is only valid for the simplified case of a single free generic particle.  Towards this end, the framework must be expanded to allow for arbitrary potentials, multiple particles, and non-zero values of spin and charge.  While these major tasks cannot be accomplished in this paper, there are promising avenues for incorporating them into this general framework that will now be discussed.

For non-zero potentials, NRQM simply adds a potential term to the Schr\"odinger equation.  In relativistic theory, this is found to be equivalent to introducing an electrical potential.  Unfortunately, when this type of potential is extended to the KGE, it is found that the positive- and negative- frequency KGE solutions respond in an opposite manner (leading to the interpretation that the negative-frequency solutions are antiparticles).  However, if one wants to compare the approach in this paper to standard NRQM, one needs \textit{all} of the KGE solutions to respond to the potential in the same manner.   General relativity provides a way to add such a scalar potential: via a metric, $g_{\mu\nu}$, in the weak-field limit.  This does not work for NRQM, because the Schr\"odinger equation is not covariant, but it should work for the KGE, which can be written $g^{\mu\nu}\nabla_\mu\nabla_\nu \phi = m^2\phi$ (with covariant derivatives).  Using this curved-space KGE to introduce potentials may be mathematically awkward, but if successful it would allow for a much easier extension to quantum gravity.  

For multiple particles, this approach could potentially allow a simplification that is not possible in NRQM: keeping the parameter space $\phi(\bm{r},t)$ fixed for any number of (identical) particles.  For two particles in standard NRQM, one expands the dimensionality of the Hilbert space, equivalent to a wavefunction $\psi(\bm{r}_1,\bm{r}_2,t)$.  This expansion is required because of the role of probability in NRQM: the wavefunction $\psi$ traditionally encodes all conditional probabilities of all possible outcomes, and the number of possible outcomes increases dramatically with the number of particles (because different types of measurements can be made on each particle).  Encoding all of these possibilities requires a wavefunction in a large configuration space.

But the picture of probability described in section 3 is quite different.  Instead of encoding all possible outcomes, the field $\phi$ is a \textit{particular solution} to the pair of preparation/measurements that actually happen -- \textit{not} all possible measurements.  To find the outcome probability for any given experimental set-up, one need only compare the permitted range of $W$ over the outcomes permitted by that particular set-up.  Instead of the wasteful increase in the parameter space of $\psi$ to deal with potential measurements that never actually happen, $\phi$ only ``bothers'' to encode the result of the \textit{actual} next measurement.  This framework therefore opens up the possibility that each type of particle might be represented by a field in physical space, and configuration space would only be a tool to summarize knowledge of this physical field.  (Such speculation agrees with the conclusion of a recent paper by Montina \cite{Montina}.)  The outcome of such a research path is far from certain, and would complicate development of a generalized measurement theory (the amplitude of the field must then be related to the number of particles present) but this also would bring quantum theory more in line with general relativity.

Finally, spin and charge will need to be added to this generic-particle framework.  Incorporating spin should be straightforward, as it is known to involve expanding the scalar field $\phi$ into a multi-component spinor.  For example, the above methodology could be extended to neutral spin-1/2 particles simply by using the second-order van der Waerden equation instead of the KGE.  Adding charge is less obvious, and might even seem problematic, as the standard interpretation of the KGE treats the $b(\bm{k})$ terms evolving like $exp(i\omega t)$ as antimatter.  Meanwhile, the above approach requires both the $a(\bm{k})$ and $b(\bm{k})$ terms to describe a single species, consistent with standard relativistic QM for spinless, neutral particles.  But doubling the dimensionality of the scalar field $\phi$ would provide parameters for both matter and antimatter while still retaining the above two-boundary formalism.  So, for example, the electron/positron field would then require a four-component (Dirac) spinor; one doubling of the scalar field to introduce antimatter, and another doubling to permit spin-1/2.  But instead of simply using the first-order Dirac equation, this framework requires a second-order differential equation.  For the electron/positron field, the most likely candidate would be the \textit{square} of the Dirac equation.  In the absence of electromagnetic coupling, this simply reduces to the KGE on a Dirac spinor, making the connection to the above results quite apparent. 

This approach also has implications for quantum foundations.  One ongoing disagreement concerns whether or not the solution to the Schr\"odinger equation $\psi$ corresponds to some observer-indepedent  ``reality'', or is instead a construct concerning our knowledge of a system.  This approach offers a middle ground; $\psi$ would be a construct, while the underlying system would be best described by some classical field $\phi$.  Without knowing the future boundary condition, the best approximation one can make concerning $\phi$ would be a guess that looks something like $\psi$, but after the fact, one can reconstruct what actually happened between measurements.  Non-classical behavior of those fields, such as violations of Bell's inequality, might now be explained by virtue of the parameters $t_0$, $a(\bm{k})$, and $b(\bm{k})$ all being hidden variables that are not ``local'' as defined by Bell \cite{Bell} (because they depend on future events).  

Such a ``realistic'' view of $\phi$ is more plausible if $\epsilon$ does not exactly equal zero, because otherwise $\phi$ has discrete infinite coefficients at $k=k_n$.  But there is some hope for a physical motivation of a non-zero epsilon; invoking a small but non-zero imaginary mass component has a history that goes back to Dirac \cite{Dirac} and has been revived several times since \cite{Nakanishi}\cite{Kleefeld}.  Another promising avenue to a real $\phi$ can be found by writing the Klein-Gordon equation in curved space, where the Christoffel symbols enter the equation in a similar manner as $\epsilon$ \cite{Evans}.  Given that no quantum mechanics experiment has been performed in an exactly flat space, it would be possible for a slightly curved space-time metric to be the source of a non-zero $\epsilon$. 

It may be considered a disadvantage to this approach that $\phi$ depends on future events.  The most serious concern is the possibility of constructing a causal paradox.  If a system $\phi$ is constrained between two temporal boundaries, then $\phi$ contains some information about the future.  At first glance, it seems logical that an experimenter could 1) measure $\phi$ before the future boundary, 2) learn about future events, and 3) change those events to something different.  This would create an irresolvable paradox, sufficient reason to end this line of inquiry.  But given the assumed equivalence between measurements and boundary conditions, step 1) is automatically forbidden; the measurement of $\phi$ \textit{is} the physical constraint corresponding to the future boundary condition.  To measure $\phi$ before the imposed constraint would be a contradiction in terms; in this framework there is no such thing as extracting information without also imposing that same information as a boundary condition. 

Indeed, there is a perspective that makes this retrocausal aspect an advantage, not a disadvantage \cite{Miller}.  It is obvious that quantum mechanics is counter-intuitive, but it must be counter-intuitive for a reason -- some human intuition that fundamentally contradicts some physical principle.  One example of this would be the well-known conflict between our directed experience of time and the more symmetric treatment of time in fundamental physics.  If the counter-intuitive aspects of quantum mechanics could be explained via classical fields symmetrically constrained by both past and future events, then it would be a mistake to reject such a solution based solely on our time-asymmetric intuitions.

Regardless, the ultimate test for any theoretical proposal is experiment.  Perhaps the strongest reason to pursue this line of research is that it leads to results which contradict standard quantum mechanics (albeit only in experimentally-difficult regimes).   Hopefully an incorporation of potentials and a relativistically correct measurement theory will lead to practical experiments which can confirm or deny the validity of this general approach.

\section*{APPENDIX}

This appendix derives the limits in which the approximation (\ref{eq:approx}) holds good.  Recall $k_n$ are the solutions to the equation $\omega_n({k_n})=n\pi /t_0$, and $n_0$ is the smallest value of $n$ for which there is a solution.  It will also be useful to define the quantity $k_{n+1/2}$ using the similar equation $\omega_n(\bm{k}_{n+1/2})=(n+1/2)\pi /t_0$.  To compare to NRQM, one is interested in the limit where $\omega\approx mc^2/\hbar + \hbar k^2/(2m)$.

Working backwards from the right side of (\ref{eq:approx}), using spherical k-coordinates, the ``radial'' component of this integral can then be written as the discrete sum of integrals
\begin{eqnarray}
\label{eq:fint}
\int^\infty_0 Re[FG^*] cos(\omega t_0) k^2 dk =&\sum_{n=n_0+1}^\infty \int^{k_{n+1/2}}_{k_{n-1/2}} \!\!\! Re[FG^*] cos(\omega t_0) k^2 dk+\nonumber\\
& \int^{k_{1/2}}_{0} \!\!\! Re[FG^*] cos(\omega t_0) k^2 dk.
\end{eqnarray}
The final integral (where $k_{n_0+1/2}$ is written as $k_{1/2}$ for clarity) is different from the others because the lower limit must be forced to zero, and the precise value of $k_{n_0-1/2}$ depends on the values of $m$ and $t_0$ -- indeed, it may not even exist.  In the special case that $k_{n_0-1/2}=0$, then $k_{n_0+1/2}=(2\pi m/\hbar t_0)^{1/2}\equiv k_0$.  This value, $k_0$, is a reasonable estimate for the value of $k_{n_0}$, given that $m$ and $t_0$ are not known precisely.

The last integral in (\ref{eq:fint}) is just one of many similar integrals that scale like $k^2$, so this lowest-k integral it is not likely to be important unless $F(k)G^*(k)$ is only large for values of $k<k_{0}$.  But such a narrow range is only possible if either $(1/F)dF/dk$ or $(1/G)dG/dk$ evaluated at $k\le k_{0}$ is on the order of $k_{0}$.  This leads to the first constraints on the approximation (\ref{eq:approx}), 
\begin{equation}
\label{eq:lim1}
\left| \frac{\partial F}{\partial k} \right|< \frac{|F(k)|}{k_0}, \,\,\,\,\, \left| \frac{\partial G}{\partial k} \right|< \frac{|G(k)|}{k_0}\,\,\,\,\, for\, k\le k_0.
\end{equation}
This constraint is discussed above in section 5 on the uncertainty principle.

Each integrals in the sum in (\ref{eq:fint}) will be referred to as $Y(k_n)$; writing out the k-dependence explicitly (in the NR limit) 
\begin{multline}
\label{eq:jint}
Y(k_n)=\int^{k_{n+1/2}}_{k_{n-1/2}} Re[F(k)G^*(k)] [cos(\omega_0 t_0)cos(\alpha^2 k^2)- \\
sin(\omega_0 t_0)sin(\alpha^2 k^2)] k^2 dk,
\end{multline}
using the constants $\omega_0=mc^2/\hbar$ and $\alpha=\sqrt{\hbar t_0/(2m)}$.

Taylor expanding the nth integrand around $g(k_n)=Re[F(k_n)G^*(k_n)]$, the zero-order term in the expansion can be integrated exactly:
\begin{multline}
\label{eq:jint2}
Y(k_n)=g(k_n)\frac{k}{2\alpha^2}sin(\omega_0t_0+\alpha^2k^2)+O[g^\prime(k_n)]- \\
\left. g(k_n)\sqrt{\frac{\pi}{8\alpha^6}}[cos(\omega_0t_0)S(\alpha k)+sin(\omega_0t_0)C(\alpha k)] \right|^{k_{n+1/2}}_{k_{n-1/2}}.
\end{multline}
Here $C(x)$ and $S(x)$ are the Fresnel integrals, which for $x>\pi$ can be excellently approximated by \cite{Fresnel}
\begin{equation}
\label{eq:Cx}
C(x)=\frac{1}{2}+\frac{1}{\sqrt{2\pi}x}sin(x^2)-\frac{1}{\sqrt{8\pi}x^3}cos(x^2)+O(x^{-5})sin(x^2)+O(x^{-7})cos(x^2),
\end{equation}
\begin{equation}
\label{eq:Sx}
S(x)=\frac{1}{2}-\frac{1}{\sqrt{2\pi}x}cos(x^2)-\frac{1}{\sqrt{8\pi}x^3}sin(x^2)-O(x^{-5})cos(x^2)+O(x^{-7})sin(x^2).
\end{equation}
With $x=\alpha k$, the constraint $x>\pi$  corresponds to $k>k_0$, reinforcing the earlier discovery that this approximation fails when the first integral in (\ref{eq:fint}) dominates the others, and therefore is equivalent to the earlier constraint (\ref{eq:lim1}).  Plugging in the limits in (\ref{eq:jint2}), dramatic simplifications occur because $sin(\omega_{n+1/2}t_0)=(-1)^n$ and $cos(\omega_{n+1/2}t_0)=0$;
\begin{multline}
\label{eq:jint3}
Y(k_n)\!= g(k_n)(-1)^n\!\!\left[\frac{(k_{n+1/2}+k_{n-1/2})}{2\alpha^2}+\frac{(k^{-3}_{n+1/2}+k^{-3}_{n-1/2})}{8\alpha^6}\right]\!+\\
O(\alpha^{-10}k^{-7})+O(g^\prime).
\end{multline}
\begin{equation}
\label{eq:jint4}
Y(k_n)\approx g(k_n)(-1)^n\frac{2mk_{n}}{\hbar t_0}+O(\alpha^{-10}k^{-7})+O(g^\prime).
\end{equation}
The last step is an approximation, good to better than 1\% even for the first term ($n=n_0+1$), and rapidly improving for the higher terms in the sum.  Within a constant, the first term in (\ref{eq:jint4}) is exactly the left side of (\ref{eq:approx}), with the correct scaling of both $k_n$ and $t_0$. 

The next term in the Taylor expansion, $g^\prime(k)=d\, Re[F(k)G^*(k)]/dk$, is unimportant if the variation with $k$ is sufficiently slow.  This integral is not much harder than the last, because the $k^3cos(k^2)$ integral can be done by parts without requiring the Fresnel integrals.  To highest surviving orders, this term becomes
\begin{multline}
\label{eq:gprime}
O(g^\prime)\!=-g^\prime(k_n)\frac{(-1)^n}{2\alpha^2}\!\left\{ k^2_{n+1/2}+k^2_{n-1/2}\right. - \\
\left. k_n\!\left[k_{n+1/2}+k_{n-1/2}+(k^{-3}_{n+1/2}+k^{-3}_{n-1/2})/(4\alpha^4)\right]\right\},
\end{multline}
\begin{equation}
\label{eq:jint5}
Y(k_n)\approx g(k_n)(-1)^n\frac{2mk_{n}}{\hbar t_0}-\frac{\pi^2-8}{32}\left(\frac{2m}{\hbar t_0}\right)^3\frac{-1^ng^\prime(k_n)}{k^2_n}+O(\alpha^{-10}k^{-7}).
\end{equation}

Again, this last step is a quite accurate numerical approximation that increases in accuracy with $n$.  Comparing the magnitude of these two terms, one finds that the approximation (\ref{eq:approx}) fails unless the fourier expansion of both boundary conditions vary sufficently slowly over k such that 
\begin{equation}
\label{eq:lim2}
\left| \frac{\partial}{\partial k} [F(k)G^*(k)]\right|_{k_n} \ll 170\,|F(k_n)G^*(k_n)|\frac{k^3_n}{k^4_0}
\end{equation}
This result is also is discussed above in section 5 on the uncertainty principle.  The conditions on the higher order terms in the Taylor expansion are not worked out here, but they should also be taken into account.

\section*{Acknowledgments}
The author is eternally grateful to J. Finkelstein for careful and thoughtful criticism.  Many additional improvements arose thanks to detailed analysis from E. Cavalcanti.  Further thanks go to M. Derakhshani, P. Goyal, F. Kleefeld, D. Miller, R. Schafer, L.S. Schulman, R.W. Spekkens, W. Struyve, and W.R. Wharton.

\end{document}